\documentclass{article}
\usepackage{spconf,amsmath,graphicx}

\usepackage{cite}
\usepackage{amsmath,amssymb,amsfonts}
\usepackage{algorithmic}
\usepackage{graphicx}
\usepackage{textcomp}
\usepackage{xcolor}
\usepackage[ruled,linesnumbered]{algorithm2e}
\usepackage{subfigure}
\usepackage{booktabs}
\usepackage{multirow}
\usepackage{array}
\usepackage{arydshln}
\usepackage{enumitem}
\usepackage{setspace}
\usepackage{listings}
\usepackage[export]
{adjustbox}

\title{Improving EEG-based Emotion Recognition by Fusing Time-frequency And Spatial Representations}
\name{Kexin Zhu$^\ddag$, Xulong Zhang$^\ddag$, Jianzong Wang$^\ast$, Ning Cheng, Jing Xiao\thanks{\ddag These authors contributed equally to this work.}\thanks{$^\ast$Corresponding author: Jianzong Wang, jzwang@188.com.}}
\address{Ping An Technology (Shenzhen) Co., Ltd., China}

\begin{document}
%
\maketitle
\begin{abstract}
Using deep learning methods to classify EEG signals can accurately identify people's emotions. However, existing studies have rarely considered the application of the information in another domain's representations to feature selection in the time-frequency domain. We propose a classification network of EEG signals based on the cross-domain feature fusion method, which makes the network more focused on the features most related to brain activities and thinking changes by using the multi-domain attention mechanism. In addition, we propose a two-step fusion method and apply these methods to the EEG emotion recognition network. Experimental results show that our proposed network, which combines multiple representations in the time-frequency domain and spatial domain, outperforms previous methods on public datasets and achieves state-of-the-art at present.
\end{abstract}
\section{Introduction}
\label{sec:intro}

Compared to the natural world, we humans know very little about ourselves. A person can interact with other people, machines, or nature in many ways~\cite{zhang2022metasid, duan2022melody}, such as sight, hearing, speech, gestures, writing, etc. As the research progresses, especially in the field of brain science, we have the opportunity to bypass these cumbersome interaction methods and interact directly through the electrical signals of the brain.

Electroencephalogram (EEG) signal is a spontaneous electrical signal generated by the conscious activity of the human brain~\cite{DBLP:journals/robotica/PoojaPV22}, and the information extracted from EEG signal can be used as an important indicator to study the conscious state of the human brain
. For example, classifying EEG signals according to certain rules can accurately identify human emotions, or control the activities of mechanical prostheses. Compared with other external manifestations such as expressions, gestures, language, etc., extracting raw information from spontaneous EEG signals is of great significance because one cannot control or mask these spontaneously generated EEG signals~\cite{DBLP:journals/taffco/ZhaoGSWW18}. In addition, it is extremely difficult for people with language barriers or physical disabilities to recognize information from sounds, gestures, etc. It can be said that EEG is one of the most suitable means to extract information about human conscious activities.

Medical research has proved that human brain activities are mainly reflected in the changes of several frequency bands of EEG signals (i.e. features of frequency domain), and these changes are related in both time series and brain region
~\cite{DBLP:conf/isbra/WangHLJCYF21}
, which are reflected in the context correlation of signal changes and correlation between electrodes at different locations. In recent years, many EEG classification models based on temporal, frequency, and spatial features have been proposed. Hou and Jia et al.~\cite{Hou2020DeepFM} applied long short-term memory (LSTM) to extract features and used graph convolution networks (GCN) to model topological structure. He et al.~\cite{DBLP:conf/icassp/HeLWS22} applied channel attention to multi-layer perceptron to adaptively learn the importance of each channel. Specific to the task of EEG emotion recognition, many researchers have also proposed some approaches. Yin et al.~\cite{DBLP:journals/asc/YinZHZC21} used GCN to extract the relationship between channels as spatial features and use LSTM to memorize the relationship changes. He and Zhong et al.~\cite{DBLP:conf/icassp/HeZP22} proposed a model that combines temporal convolution networks with adversarial discriminative domain adaptation to solve the problem of domain drift in the cross-subject emotion recognition task. Some of the existing works focus on the representations of different domains, lacking the mapping process of features between representations, and some fusion methods are difficult to combine different levels of feature information to comprehensively model EEG signals.

\begin{figure*}[htb]
\begin{minipage}[b]{1.0\linewidth}
  \centering
  \centerline{\includegraphics[width=\linewidth]{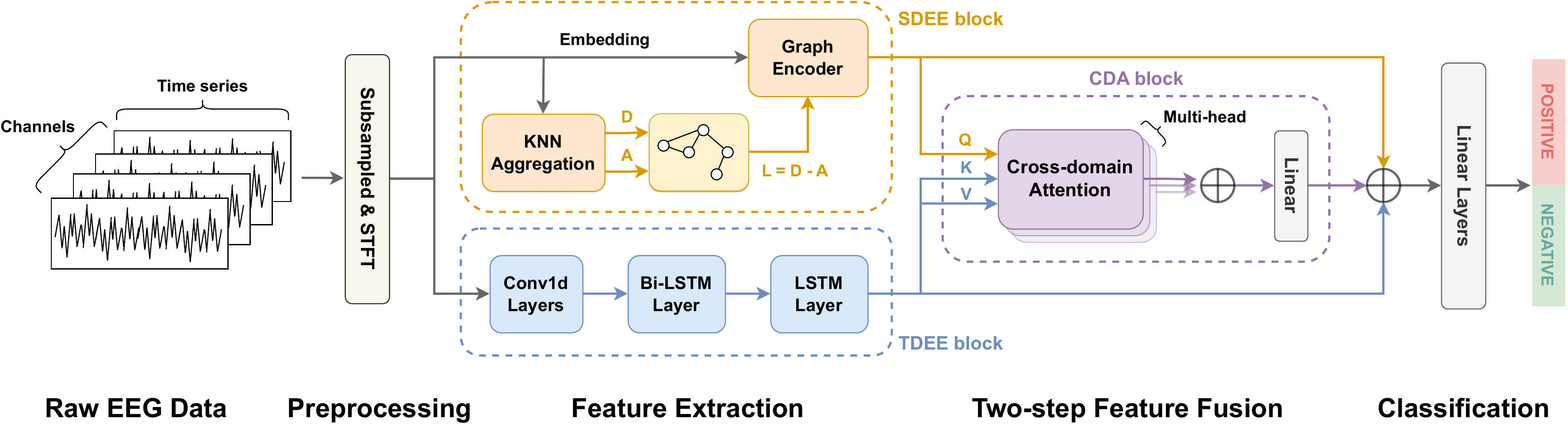}}
\end{minipage}
\caption{Overall architecture. Our proposed model takes multi-channel EEG time series data as input and outputs binary prediction results. During the two-step fusion process, the CDA block completes the first fusion, and then the feature vectors output by each block are concatenated to complete the second fusion.}
\label{fig:res}
\end{figure*}

In this paper, we propose a multi-domain feature fusion model for emotion recognition based on EEG, which fuses features from multiple domains such as temporal, frequency, and spatial representations.
In particular, we propose a two-step fusion method to improve the multi-domain feature fusion process, to better preserve the respective information of each domain. Different from other methods in the same task, we believe that the application of the multi-domain fusion method can introduce the connection between each channel obtained by the graph neural network to select features in the time-frequency feature representations, and better help the model to extract correlations in time-frequency features.
We conducted rigorous experiments, including the comparison with other existing methods, the selection of graph encoder, and the ablation experiment of each block in the network. According to experimental results, our proposed method has achieved the highest accuracy in user-independent emotion recognition.

Our main contributions are listed below:
\vspace{-4pt}
\begin{enumerate}
\setlength{\abovecaptionskip}{0pt}
\setlength{\itemsep}{1pt}
\setlength{\parsep}{1pt}
\setlength{\parskip}{1pt}
\item[1)]
We proposed a method for EEG multi-domain feature fusion using cross-domain attention, which utilizes information from spatial representations to assist in the selection of time-frequency features.
\item[2)]
Based on applying the multi-domain method for feature fusion, we proposed a two-step fusion method to preserve more feature information in time-frequency and spatial feature vectors.
\item[3)]
We applied described method in the proposed emotion recognition network. We performed experiments and the results show that our proposed method has a better effect on EEG-based emotion recognition and achieves the best performance.
\end{enumerate}
\vspace{-4pt}
\section{Method}
The overview is illustrated in Figure \ref{fig:res}. Our network is mainly composed of three parts, namely, a time-domain embedding {\&} encoding block (TDEE block), a spatial-domain embedding {\&} encoding block (SDEE block), and a cross-domain attention block (CDA block). The processing flow of the model is as follows: after pre-processing, EEG signals are sent to the TDEE block and SDEE block to construct time-frequency domain and spatial domain features. CDA block fuses these features in the first step, and the outputs of our proposed three blocks are concatenated to complete the second step of fusion. Finally, multiple linear layers complete the classification task of emotion recognition. During the training process, the parameters in each block are updated by back-propagation.





\subsection{TDEE Block}


Our proposed TDEE block has a similar structure to CRNN~\cite{DBLP:journals/pami/ShiBY17}. It can preliminarily extract channel correlation information by convolution operation between channels and encode the context relationship by LSTM layers.

The block contains three one-dimensional convolution layers and several LSTM layers. The first two convolutional layers are followed by a ReLU, and the last layer is followed by a batch normalization layer.
We improved the original CRNN and added an LSTM layer after the Bidirectional LSTM layer to enhance the ability of the network to extract temporal context information.

\subsection{SDEE Block}

Our proposed SDEE block contains an embedding layer, a graph encoder, and a series of processing steps for constructing graphs. The multi-channel EEG signal is constructed as a graph structure with each channel as nodes, and the adjacency matrix $A$ and degree matrix $D$ are obtained by using K nearest neighbors (KNN) to calculate the connection relationship between nodes
. 
The Laplacian matrix $L$ is calculated by $L=D-A$ and used in the following graph encoder.

Many graph encoders have been proposed in previous studies, including GCN~\cite{DBLP:conf/embc/DemirKWHE21}
and GAT~\cite{DBLP:conf/iclr/VelickovicCCRLB18}. In each layer of GCN, nodes of the graph acquire information through the adjacent or further nodes. Suppose $i$ is a graph node, $N_i$ is the set of all neighboring nodes of node $i$, and $l$ is the current GCN layer. The feature of $l+1$ layer nodes can be calculated as follows:
\begin{equation}
\label{form2}
    h_{i}^{l+1}=\sigma(\sum_{j}\frac{h_{i}^{l}W^{l}}{c_{ij}})
\end{equation}
\begin{equation}
\label{form3}
    c_{ij}=\sqrt{d_{i}d_{j}}
\end{equation}
where $j$ is one of the neighboring nodes of node $i$, $d_{i}$ and $d_{j}$ are the degrees of nodes $i$ and $j$ respectively. $h_{i}^{l}$ is feature vector of node $i$ in $l$ layer. GAT improves the normalization constant in the GCN layer into the neighbor node feature aggregation function using attention weight:
\begin{equation}
\label{form4}
    h_{i}^{l+1}=\sigma(\sum_{j}\alpha_{ij}^{l}z_{j}^{l})
\end{equation}
\begin{equation}
\label{form5}
    \alpha_{ij}^{l}=\frac{exp(e_{ij}^{l})}{\sum_{k}exp(e_{ik}^{l})}
\end{equation}
\begin{equation}
\label{form6}
    e_{ij}^{l}=LeakyReLU((\vec{a}^{l})^{T}(z_{i}^{l}||z_{j}^{l}))
\end{equation}
\begin{equation}
\label{form7}
    z_{i}^{l}=h_{i}^{l}W^{l}
\end{equation}
$W^{l}$ is a trainable parameter matrix, $k\in{N_i}$. Formulas (\ref{form5}) and (\ref{form6}) respectively calculate the attention weight of nodes and the attention between pairs of nodes. 
We have applied different graph encoders for comparative experiments to determine the best graph encoding approach.

\subsection{CDA Block}

In the CDA block, we introduce multi-head cross-domain attention to applying the graph representation information of channels to feature selection, so that the network can focus on the features most related to human emotions. Cross-domain attention is inspired by cross-modal attention~\cite{DBLP:conf/acl/TsaiBLKMS19, DBLP:conf/icmcs/ZhaoRYWLL22}
, it is also a mapping relationship similar to self-attention~\cite{DBLP:conf/nips/VaswaniSPUJGKP17}, but query and key-value pairs come from different domain representations respectively. 

Let $\alpha$ represents the domain of the SDEE block output feature vector (i.e. spacial domain), and $\beta$ represents the domain of the TDEE block output feature vector (i.e. time-frequency domain). The query $Q_{\alpha}$ of each attention head can be obtained by linear mapping:
\begin{equation}
\label{form8}
	Q_{\alpha}= X_{\alpha}W_{Q}
\end{equation}
where $X_{\alpha}$ is the input with domain $\alpha$, and $W_{Q}$ is a trainable parameter matrix. Key $K$ and value $V$ are calculated in a similar way to the query:
\begin{equation}
\label{form9}
    K_{\beta}= X_{\beta}W_{K}
\end{equation}
\begin{equation}
\label{form10}
    V_{\beta}= X_{\beta}W_{V}
\end{equation}
except that their input comes from domain $\beta$. Each attention head is calculated separately:
\begin{equation}
\label{form11}
    Attention(Q_{\alpha}, K_{\beta}, V_{\beta})= Softmax(\frac{Q_{\alpha}(K_{\beta})^T}{\sqrt{d}})V_{\beta}
\end{equation}
where $(\cdot)^T$ is matrix transposition, $d$ represents the dimension of $K$. The attention of each head is concatenated and multiplied by the weight matrix to get multi-head attention:
\begin{equation}
\label{form12}
\begin{split}
    head_{i} & = Attention(Q_{\alpha}^{i},K_{\beta}^{i},V_{\beta}^{i})\\
    & =Softmax(\frac{Q_{\alpha}^{i}(K_{\beta}^{i})^T}{\sqrt{d}})V_{\beta}^{i}\\
    & =Softmax(\frac{X_{\alpha}W_{Q}^{i}(X_{\beta}W_{K}^{i})^T}{\sqrt{d}})X_{\beta}W_{V}^{i}
\end{split}
\end{equation}
\begin{equation}
\label{form13}
    MultiHead(X_{\alpha},X_{\beta})= Concat(head_{1},...,head_{H})W_{O}
\end{equation}
where $1\leq i\leq{H}$, hyperparameter $H$ is the number of attention heads, $W_{O}$ is a parameter matrix. These trainable weight matrices enhance the fitting ability of the proposed model. Multi-head attention provides different representation subspaces to learn different features, while input data from different domains provide different features, which are fused by the attention mechanism.

\subsection{Two-step Feature Fusion}
By applying cross-domain attention, our proposed model can introduce spatial topological information extracted from graph networks and pay more attention to the channels (i.e. features output by TDEE block) most related to brain emotional changes. However, the information contained in the feature vectors obtained by the graph encoder and the time series encoder is more than that. For example, the change of spatial feature vector or time-frequency feature vector itself can reflect the change in brain electrical signal when a certain emotion is generated to a certain extent.

We propose a two-step feature fusion strategy to verify this hypothesis: in the first step, the feature vectors output by the SDEE block and TDEE block are fused in the CDA block and transformed into fusion vectors, as described in the previous subsection. In the second step, the fused vector and the feature vector before fusion are concatenated and then sent to the dense layers to complete emotion classification:
\begin{equation}
\label{form14}
    X_{CM}=MultiHead(X_{\alpha},X_{\beta})
\end{equation}
\begin{equation}
\label{form15}
    X_{FC}=Concat(X_{\alpha},X_{\beta},X_{CM})
\end{equation}
where $X_{FC}$ is the input of the classifier, $X_{\alpha}$ and $X_{\beta}$ are features extracted by the SDEE block and TDEE block respectively, and $X_{CM}$ is the cross-domain fusion feature calculated by CDA block in the first step.


\section{EXPERIMENTS AND RESULTS}
\label{sec:results}

\subsection{Data Preparations}
We use the open source dataset DEAP~\cite{DBLP:journals/taffco/KoelstraMSLYEPNP12} to evaluate our proposed model. DEAP dataset consists of EEG data of 32 participants, including equal numbers of male and female subjects. Each participant watched 40 videos of about one minute and EEG signals of the participants were recorded during the watching process. After watching, each participant scored  \textit{Valence}, \textit{Arousal}, and other measures of the videos. The collected raw EEG data has been subjected to many pre-processing steps, including down-sampling and eliminating the electrooculogram.

Before the experiment, we used some other pre-processing methods for the DEAP dataset. We divide the time-series EEG signals into time slices by using a sliding window, which keeps the time-series context information to the maximum extent. Short-time Fourier transform (STFT) is used to calculate differential entropy (DE) as the frequency domain features in five bands.


\subsection{Experimental Settings}

Many experiments are conducted to determine the hyperparameters that can make our proposed model work best. Adam with a learning rate of 0.001 is used as the optimizer and cross-entropy is used as the loss function. The number of attention heads $H$ is set to 8. The sliding window has a width of 2 seconds and moves in steps of 0.125 seconds.

To confirm our experimental results, we used leave-one-subject-out (LOSO) cross-validation method.
In each validation, one subject's data is not involved in the training process but used as test data, thus realizing user-independent emotion recognition.

\subsection{Results and Discussion}

\begin{table}[]
\caption{Comparison between our proposed method and other methods, where T-F represents the time-frequency feature.}
\label{tab:1}
\begin{center}
\begin{tabular}{@{}cccc@{}}
\toprule
\multirow{2}{*}{Study} & \multirow{2}{*}{Feature(s)} & \multicolumn{2}{c}{Accuracy} \\ \cmidrule(l){3-4} 
                       &                             & Valence       & Arousal      \\ \midrule
Li et al.~\cite{DBLP:conf/icassp/LiCZCS21}            & T-F                          & 0.691         & 0.710        \\
Wang et al.~\cite{DBLP:conf/isbra/WangHLJCYF21}              & SFM                 & 0.712         & 0.713        \\
Atkinson et al.~\cite{DBLP:journals/eswa/AtkinsonC16}            & mRMR                          & 0.731         & 0.730        \\
Guo et al.~\cite{0Multi}        & T-F, FuzzyEn                          &0.844               &0.856              \\ \noalign{\vskip 1mm}\cdashline{1-4}[1pt/1pt]\noalign{\vskip 1mm}
Ours (with GAT)              & T-F, Graph                          &0.859               &0.878              \\\textbf{Ours (with GCN)}              & \textbf{T-F, Graph}                           & \textbf{0.861}         & \textbf{0.884}        \\ \bottomrule
\end{tabular}
\end{center}
\end{table}
We conducted three main tasks to evaluate the performance of our proposed model. Comparative experiments are performed to verify the improvement of classification accuracy of our proposed model compared with other approaches. GCN and GAT are applied for performance comparison, to determine the best graph encoding approach. The ablation experiments of several blocks are performed to verify the functions of each block and the feature fusion approach in our proposed model.

In the first two experiments, we compared the results of our proposed method with previous works. Table \ref{tab:1} shows the experimental results of existing methods and our proposed methods, in which Ours (with GCN) and Ours (with GAT) represent the experimental results when GCN or GAT are used as graph encoders, respectively. Experimental results have proved our proposed method achieves the best in the classification task of emotion recognition and surpasses previous works. The accuracy is close when GCN and GAT are used as graph encoders, but GAT has higher computational efficiency because of its different mechanism compared with GCN.

In the third experiment shown in Table \ref{tab:2}, we have done the ablation experiment for each block in the proposed method and respectively experimented with one-step and two-step fusion methods. Results show that our proposed multi-domain feature fusion method can effectively improve the classification accuracy, both in \textit{Valence} and \textit{Arousal} dimensions, and the effect of two-step fusion is better than that of one-step fusion.

It is worth noting that the channels that the CDA block makes the model focus on do not refer to the specific original EEG input channels, because these channels have undergone one-dimensional convolution in the TDEE block and generated new feature representations, which contain the preliminary relationship between channels and information of different brain regions. Besides, our proposed method can be used not only for emotion recognition tasks but also for other tasks related to EEG signals, as other tasks may use similar EEG bands and features.

\begin{table}[]
\caption{Ablation study of blocks in proposed method and comparison experiment of different fusion methods.}
\label{tab:2}
\begin{center}
\begin{tabular}{@{}cccccc@{}}
\toprule
\multirow{2}{*}{SDEE} & \multirow{2}{*}{TDEE} & \multirow{2}{*}{CDA} & \multirow{2}{*}{Fusion} & \multicolumn{2}{c}{Accuracy}      \\ \cmidrule(l){5-6} 
                      &                       &                      &                         & Valence         & Arousal         \\ \midrule
\checkmark            &                       &                      & -                       & 0.530          & 0.512          \\
                      & \checkmark            &                      & -                       & 0.834          & 0.840          \\
\checkmark            & \checkmark            &                      & Concat                  & 0.849          & 0.864          \\ \noalign{\vskip 1mm}\cdashline{1-6}[1pt/1pt]\noalign{\vskip 1mm}
\checkmark            & \checkmark            & \checkmark           & One-step                & 0.855          & 0.867          \\
\checkmark            & \checkmark            & \checkmark           & \textbf{Two-step}                & \textbf{0.861} & \textbf{0.884} \\ \bottomrule
\end{tabular}
\end{center}
\end{table}

\section{CONCLUSION}
\label{sec:conc}
We proposed a multi-domain feature fusion model for EEG-based emotion recognition. Specifically, we use a cross-domain feature fusion method to combine the spatial domain information to make the model focus on the time-frequency domain features most related to our task, and further combine these features by using a two-step fusion method. Experiments show the effectiveness of the proposed model. We will explore the feasibility of applying federated learning methods to deploy the model on distributed devices and use it to help more people.
%


\section{Acknowledgement}
Supported by the Key Research and Development Program of Guangdong Province (grant No. 2021B0101400003) and Corresponding author is Jianzong Wang (jzwang@188.com).

\vfill\pagebreak


\bibliographystyle{IEEEbib}
\bibliography{my}

\end{document}